# On the Use of Non-Linear Metasurfaces for Circumventing Fundamental Limits of Mantle Cloaking for Antennas

Stefano Vellucci, *Student Member*, *IEEE*, Alessio Monti, *Senior Member*, *IEEE*, Mirko Barbuto, *Senior Member*, *IEEE*, Giacomo Oliveri, *Senior Member*, *IEEE*, Marco Salucci, *Member, IEEE* Alessandro Toscano, *Senior Member*, *IEEE*, and Filiberto Bilotti, *Fellow*, *IEEE*

*Abstract*— The aim of this Communication is to investigate and demonstrate the possibility to overcome the fundamental limitations of mantle cloaking for antennas by exploiting non-linear metasurfaces. First, we recap and give additional physical insights about the fundamental bounds that apply to the electric properties of an antenna that is made invisible at its own resonance frequency. Then, an innovative strategy is proposed to circumvent these limits through the introduction of non-linear elements that are able to dynamically transform the geometry of the cloaking metasurface depending on the power level of the impinging field. Different non-linear designs are discussed, and their effectiveness is assessed through the relevant benchmark example of a half-wavelength dipole antenna able to efficiently transmit high-power signals while being invisible to low power ones. In addition, the capability of such non-linear mantle cloaks to enable the shaping of the radiation pattern of an antenna array depending on the power level of the received/transmitted signal is demonstrated. These innovative cloaking devices may find applications in different radiating systems enabling unprecedented functionalities.

## I. INTRODUCTION

In the context of electromagnetic scattering theory, the *optical theorem* relates the forward scattering amplitude of a passive object to its total extinction cross section [1]. This theorem directly implies that any sensing device able to partially absorb an impinging signal necessarily introduces some "shadows", which perturb the measurement and make the sensor detectable by an external observer. In this framework, the research has been focused on circumventing this fundamental limitation, starting from the concept of minimum-scattering antennas [2],[3]. More recently, the advent of metamaterials and cloaking techniques has given new life to this research area, specifically in view of its application towards sensor invisibility [4]-[6]. However, any sensor, even if covered with a cloaking structure, is still a passive object that must satisfy the *optical theorem*. Therefore, a cloaked passive sensor is always the result of a trade-off between the desired absorption level and its unavoidable "signature" in terms of scattered power [7].

The design of cloaked sensors has been thoroughly investigated in the last years, especially for antenna systems working at microwaves and radio frequencies. In particular, the mantle cloaking approach [8]-[10] has emerged as one of the most versatile among the available techniques suitable for antenna scenarios, since it enables the design of conformal cloaking devices that do not electromagnetically isolate the object to hide [11]-[14]. Several configurations have been proposed to reciprocally conceal antennas working at different frequencies, such that two radiating elements can be placed in close proximity without causing a mutual blockage [15]-[22]. However, the performance of such designs is not bounded by the *optical theorem*, since the cloaking structure applied to each antenna is designed to reduce its scattered power at the operating frequency of the other elements, i.e., it introduces an out-of-band scattering reduction. Therefore, the cloaked antenna actually scatters a non-negligible amount of power at its own operating frequency (i.e, *in-band*) thus allowing an efficient operation in reception/transmission.

On the contrary, the synthesis of antennas/sensors able to achieve invisibility at their own operating frequency turns out to be a much more challenging problem both from the theoretical and from the practical viewpoints. Indeed, the design of antennas able to efficiently transmit power while being invisible to an external radiating system operating at the same frequency is theoretically forbidden (as stated by the *optical theorem*) if a linear, passive, and time-invariant cloak is used. In other terms, if we cloak an antenna in the receiving mode, the presence of the cloaking structure irreparably worsens the performance of the antenna in the transmission mode.

In recent years, some attempts to circumvent the limitations of linear, passive, and time-invariant cloaking devices have been made. More specifically, the use of active non-Foster circuits has been proposed to dramatically increase the operating bandwidth of mantle cloaks [23]. Moreover, the peculiar features of parity-time symmetric systems have been proposed as a possible route to compensate losses and, thus, to allow the design of a truly invisible sensor, i.e. exhibiting zero scattering but still being able to receive the impinging signal [24],[25]. Although active circuits potentially enable to overcome the fundamental limits of passive and linear cloaking structures, they are still affected by some drawbacks, such as stability issues and the need for an external feeding structure. However, some of these limitations could be also circumvented by breaking a different assumption, i.e., the linearity of the overall system. In this regard, non-linear elements have been used to obtain a cloaking structure exhibiting a power-dependent behavior [26]-[28]. However, in these cases, the non-linear cloak is still designed for passive scatterers, and, to the best of the authors' knowledge, no efforts have

Manuscript received March XX, 2020; accepted XXX XX, 202X. Date of publication XXXX XX, 202X; date of current version XXXX XX, 202X. This work has been developed in the frame of the activities of the research contract MANTLES, funded by the Italian Ministry of Education, University and Research as a PRIN 2017 project (protocol number 2017BHFZKH). (Corresponding author: S. Vellucci).

S. Vellucci, G. Oliveri, and M. Salucci are with the ELEDIA Research Center (ELEDIA@UniTN - University of Trento), Trento, 038123 Italy (e-mail: stefano.vellucci@unitn.com).

A. Monti and M. Barbuto are with the Niccolò Cusano University, 00166, Rome, Italy.

A. Toscano, and F. Bilotti are with the Department of Engineering, ROMA TRE University, 00146 Rome, Italy.

Color versions of one or more of the figures in this communication are available online at http://ieeexplore.ieee.org.

Digital Object Identifier X

been carried out to exploit non-linearity to design an invisible antenna/sensor. In this framework, a different approach based on the use of cloaking devices able to distinguish between different waveforms has been recently proposed [29]. Thanks to the adoption of so-called waveform-selective circuits [30], an antenna has been made invisible/visible to either a short pulse or a continuous signal. However, in [29] the non-linear elements of the circuit (i.e., the diodes) are used to rectify to DC the incoming signal, and the power-dependent behavior is not properly engineered and exploited.

Inspired by these studies, the aim of this contribution is to investigate the potentialities offered by non-linear elements, such as PIN diodes, to design a covering structure allowing to cloak a radiating element at its own operating frequency. In particular, by exploiting the mantle cloaking technique, the possibility to design an antenna covered with a non-linear metasurface that exhibits minimum scattering response for low-power (LP) levels and, simultaneously, efficient radiation for high-power (HP) signals at the same frequency is demonstrated. The adoption/customization of such an approach in those applicative scenarios where a considerable difference between transmitted and received power levels (e.g., an antenna operating in a hostile environment requiring radar invisibility but normal operation in transmission) is discussed.

Furthermore, the effectiveness of non-linear cloaks is also demonstrated in invisibility problems involving multi-antenna complex systems. More in detail, the design of a bi-dimensional array of dipole antennas exhibiting a steerable high-gain beam in the transmitting mode and an omnidirectional radiation pattern for low-power received signals is illustrated [31]. It is worth noticing that, compared to [28], here we deal with the design of a non-linear cloak for the *in-band* scattering reduction of active elements. Thus, since the fundamental limitations due to the optical theorem apply, the design process of the cloak proposed here is significantly different and further complicated, as the radiating and electrical characteristics of the antenna elements must be preserved.

## II. NON-LINEAR METASURFACES FOR CLOAKED ANTENNAS

The optical (or "forward scattering") theorem defines the fundamental bounds relating the total scattered and absorbed powers to the forward scattered field of any passive scatterer [1] included, thus, sensors and antennas. From an antenna point of view, it follows that fundamental limitations exist for the absorption efficiency of radiating devices, which is defined as the ratio between the absorbed power over the sum of absorbed and scattered powers.

As known, the power scattered by an antenna is due to the combination of two different quantities: the antenna re-radiation scattering, which is due to the mismatch between the antenna input impedance and its load, and the antenna structural scattering (also referred to as residual scattering [32],[33]). The first term introduces a fundamental difference between the antennas' scattering and the scattering from passive objects. In this latter case, in fact, only structural scattering is present [32].

In the case of a matched antenna (*i.e.,* assuming zero reradiation scattering), lower and upper scattering power bounds have been derived [33], showing that an increasing directivity of the scattered power (tending to infinity) is required to maximize the absorption efficiency and, at the same time, massively reduce the structural scattering. Similar bounds have been defined for the particular case of an antenna/sensor coated by properly designed metamaterial covers [7]. In this case, however, the presence of the re-radiation scattering contribution, due to the mismatch between the antenna and its load, was considered. Indeed, it was shown that covering an antenna with a plasmonic or a mantle cloak able to minimize the antenna structural scattering (*i.e.,* designed to reduce the scattering from the antenna under open-circuit conditions) allows tuning its absorption efficiency by simply acting on the value of the load [7]. This example suggested the possibility to exploit plasmonic or mantle cloaks not only to achieve an invisibility effect minimizing the total scattering from the antenna, but also maximizing the absorption efficiency characteristic by acting just on the scattering structural term.

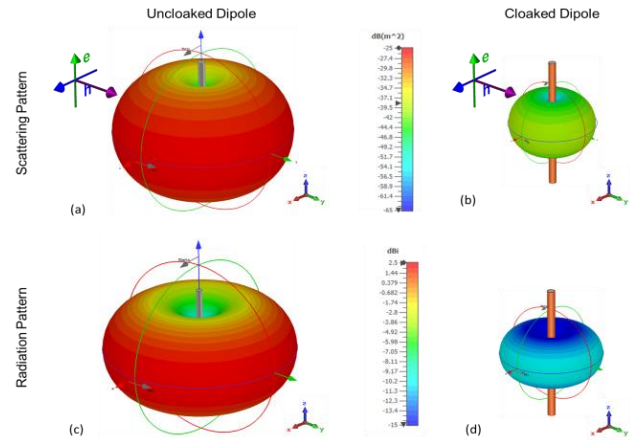

Fig. 1. (a) 3D scattering pattern and (c) 3D realized gain pattern of a half-wavelength dipole antenna on a 50 Ω load, evaluated at the resonant frequency $f_0$. (b), (d) Same as (a), (c) when the antenna is coated by a mantle cloak operating at $f_0$.

However, it is worth emphasizing that, in antenna scenarios, the presence of both the re-radiation and structural scattering terms makes the cloak design more challenging compared to the case of passive objects. In fact, the cloaking of a matched antenna introduces a perturbation of its matching condition, since the total scattering of the antenna also depends on the impedance matching through the reradiation scattering contribution.

For the sake of clarity, this phenomenon is illustrated in Fig. 1 for the general case of a half-wavelength dipole antenna resonating at the frequency $f_0$ = 3GHz. When the antenna is coated by a mantle cloak designed to suppress the scattering of the antenna at $f_0$ (Fig. 1 (b)), the realized gain pattern of the antenna (Fig. 1(d)) strongly deteriorates in comparison with the uncloaked case (Fig. 1(c)), due to the mismatch of the antenna with its load. In other terms, the cloaking device inevitably modifies the value of the antenna input impedance. As can be appreciated in Fig. 2, the imaginary part of the antenna input impedance is almost zero in the uncloaked case, and its real part turns out to be around the standard 50 Ω value. Therefore, the dipole antenna presents a good impedance matching at $f_0$. Conversely, in the cloaked case, i.e., when the antenna scattering cross section (*SCS*) is suppressed at $f_0$, the real part of the input impedance significantly increases, while the imaginary part tends to infinity (i.e., open-circuit condition), preventing the use of this cloaked antenna as an effective transmitting radiator. Thus, it is evident that, compared to the cloaking of passive scatterers, a different design approach should be used to preserve the electric and radiative characteristics of the antenna when the cloak is applied.

The idea explored in this paper is to circumvent this fundamental bound by exploiting a non-linear cloaking device able to turn ON/OFF its cloaking effects depending on the power level of the impinging wave [31]. To better explain this idea, let us consider a conventional mantle cloak for a wire antenna made of a purely reactive metasurface printed on a dielectric substrate [17]-[22].

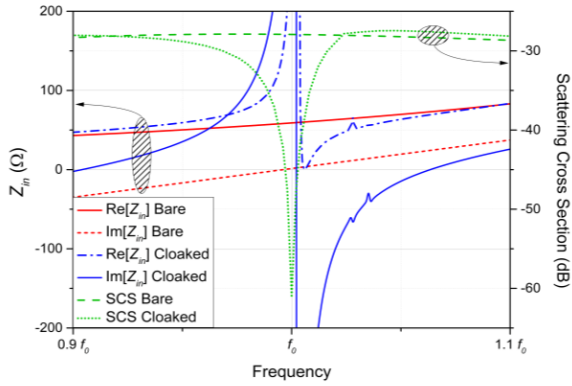

Fig. 2. Complex input impedance and total scattering cross section of a half-wavelength dipole antenna when coated by a mantle cloak operating at $f_0$ (cloaked) and w/o the coating cloak (bare).

In order to achieve a cloaking effect, reactive metasurfaces made of either vertical or horizontal sub-wavelength metallic strips can be used [12]. These structures can be effectively characterized by using a purely imaginary surface impedance ($Z_s^{cloaking}$), whose value can be analytically derived and simply controlled by acting on the width and distance between the strips [12].

To design a power-dependent mantle cloak, the gaps of these conventional metasurfaces can be loaded with diode pairs, as schematically reported in Fig. 3, where two different configurations for capacitive or inductive metasurfaces are shown. Since the equivalent impedance of the diodes switches from one of a small resistance for HP signals ($R_{HP}$), to a parallel between a large resistance ($R_{LP}$) and a capacitance ($C_{LP}$) for LP signals [34], the diodes act as non-linear elements able to short/open circuiting consecutive strips. Thus, the loaded metasurfaces exhibit an equivalent surface reactance that depends on the power level of the impinging signal.

Furthermore, as long as the power is within the operating regime of the diode, multiple switching can be obtained without incurring in bistability or irreversible effects [35]-[36].

Fig. 3 reports two different unit-cells that can be used to design either a capacitive or inductive non-linear metasurface for cloaking application. In particular, when a non-linear cloaking capacitive metasurface is needed, the configuration reported in Fig. 3(a) can be exploited. Here, the equivalent surface reactance of the metasurface is determined by the period $D$ of the metasurface, the distance $g$ between the strips, and by the impedance of the diode loading the gap between consecutive strips. It is worth noticing that the absolute value of the surface reactance of the metasurface decreases by reducing the dimension of $D$ (once fixed $g$) [12], and that the diode acts as a short-circuit in the HP scenario or as an open-circuit in the LP scenario [28]. Thus, the open/short-circuiting effect of the diodes is exploited to dynamically modify the period $D$ of the metasurface unit cell.

Specifically, in the LP scenario (Fig. 3(b)), due to the open-circuiting of the diodes, the period of the metasurface unit cell is $D = D_{LP}$ and the value of the surface reactance is $Z_s^{C*}$. Conversely, in the HP scenario (Fig. 3 (c)), the loaded strips are short-circuited and the periodicity $D$ of the unit-cell is transformed into $D_{HP}$. Thus, the value of the surface impedance is increased and is equal to $Z_s^{C**}$. Therefore, to design a capacitive metasurface whose cloaking effect turns ON for LP signals, its geometry should be chosen so that $Z_s^{C*}$ is equal to the desired cloaking value (i.e., $Z_s^{C*} = Z_s^{cloaking}$). The cloaking effect is, then, turned OFF for HP signals since the value of the surface impedance is varied to $Z_s^{C**}$. The opposite behavior (i.e., cloak ON for HP signals, OFF for LP signals) can be achieved by engineering the metasurface geometry such that $Z_s^{C**} = Z_s^{cloaking}$.

The same approach can be exploited in the case of inductive metasurface using the configuration reported in Fig. 3(d), made of continuous vertical metallic strips interleaved by truncated loaded strips. In this case, the metasurface surface reactance depends on the period $d$ and the width $w$ of the strips. In particular, the surface reactance of the metasurface decreases by reducing the dimension of $d$ for a constant value of $w$ [12]. Thus, in the LP scenario (Fig. 3(e)), due to the open-circuiting of the diodes, the period of the metasurface unit cell is $d = d_{LP}$ and a desired value of the surface reactance can be obtained ($Z_s^{L*}$); in the HP scenario (Fig. 3(f)), the loaded strip is short-circuited, the period transformed into $d = d_{HP}$, and the surface reactance is increased to $Z_s^{L**}$. Thus, to design an inductive metasurface whose cloaking effect turns ON for LP signals, its geometry should be engineered such that $Z_s^{L*} = Z_s^{cloaking}$. Instead, when an inductive metasurface whose cloaking effect turns ON for HP signals the metasurface should be designed such that its value of the surface impedance is $Z_s^{L**} = Z_s^{cloaking}$.

Finally, it is worth noticing that in the HP scenario, for both capacitive (Fig. 3(c)) and inductive configurations (Fig. 3(f)), in comparison with its equivalent unloaded value, the metasurface reactance exhibits a slight deviation due to the parasitic capacitance $C_{LP}$ of the diodes.

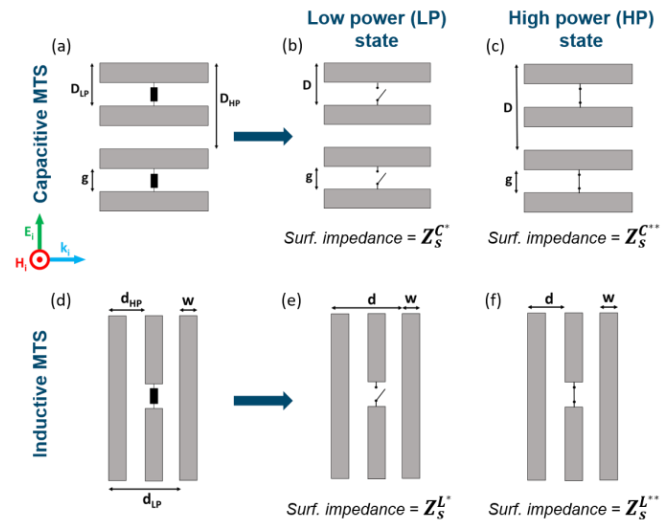

Fig. 3. Capacitive ((a), (b), (c)) or inductive ((d), (e), (f)) metasurface unit-cells loaded by diode pairs for designing non-linear mantle cloaks. (b) Equivalent schematic configuration for the capacitive metasurface unit cell for low-power (LP) signals. (c) Same as (b), but for high-power (HP) signals. (e), (f) Same as (b),(c), respectively, but for an inductive metasurface unit cell.

TABLE I

DIMENSIONS OF THE NON-LINEAR CLOAKED ANTENNA

| Symbol | Quantity | Value |
| --- | --- | --- |
| $w$ | Width of the strip | $\lambda_0 / 165$ |
| $d_{HP}$ | Strip periodicity | $\lambda_0 / 127$ |
| gap | Gap between the interleaved strip | $\lambda_0 / 100$ |
| $L$ | Length of the dipole | $\lambda_0 / 2.5$ |
| $a$ | Radius of the dipole | $\lambda_0 / 50$ |
| $g$ | Gap of the dipole | $\lambda_0 / 100$ |
| $a_c$ | Radius of the cloak | $1.2 \cdot a$ |

## III. NUMERICAL RESULTS

To demonstrate the validity of our approach, a half-wavelength dipole antenna covered with a non-linear metasurface has been designed to exhibit, at the resonant frequency $f_0$ = 3GHz, minimum scattering for LP and efficient radiation for HP signals. Since an inductive metasurface is required to cloak a dipole antenna at its resonance frequency [7], the mantle cloak configuration reported in Fig. 3(d) has been exploited, and the value of its surface impedance $Z_s^{L*}$ has been engineered to be equal to the required $Z_s^{cloaking}$ [9] able minimizing the scattering cross section of the antenna when the loading diodes are open-circuited. The final geometry is reported in Fig. 4(a), and the electrical dimensions are summarized in Table I. In particular, the cloaking metasurface consists of four vertical metallic strips that extend along the whole dipole. These strips are interleaved by a sub-array of three truncated strips. The gaps between them are loaded with a commercial diode (Hitachi HVM14S), which adds the desired power-dependent behavior to the cloaking metasurface. The metasurface is printed on a dielectric cylindrical shell of Preperm L700HF ($\varepsilon_r$ = 7.2, $\delta_{tan}$ = 0.0006). It is worth noticing that compared to Fig. 3(d), the number of truncated interleaved strips has been increased (from one to three) to emphasize the difference between the value of $Z_s^{L*}$ and $Z_s^{L**}$, i.e., when short/open circuiting the diode. Moreover, a feasible width of the strips has been also considered in the design process, and the truncated strips have been electrically connected (inset of Fig. 3(a)) to minimize the number of diodes.

To analyze the scattering and matching properties of the antenna, circuital-electromagnetic co-simulations of the loaded/unloaded device have been performed using the commercial software CST Studio Suite [37]. In particular, the total scattered power of the antenna and the magnitude of its reflection coefficient for different scenarios have been reported in Fig. 4 (b) and Fig. 4 (c).

As it can be appreciated, when the dipole is coated by the designed inductive metasurface without the loading diodes (magenta dash-dotted lines), the total scattering cross section of the antenna is massively reduced at $f_0$ compared to the uncoated case (Fig. 4(b)).

However, at the same time, the antenna is also totally mismatched at $f_0$ (Fig. 4(c)). When the cloak is loaded by the diode pairs, it is possible to appreciate the dynamic behavior of the non-linear cloak for different values of the antenna input power. In the LP scenario, for an input power lower than 0 *dBm* (continuous blue lines), a slight shift of the cloaking resonance appears due to the parasitic capacitance $C_{LP}$ exhibited by the diodes (Fig. 4(b)), and the antenna is still mismatched at $f_0$ (Fig. 4(c)). A dynamic transition can be appreciated by gradually increasing the input power: the cloaking resonance shifts towards lower frequencies progressively restoring the scattering property of the antenna at $f_0$ and, at the same time, the matching characteristic of the antenna at $f_0$ improves. Finally, in the HP scenario, i.e, for an input power higher than 40 *dBm* (red continuous lines), the cloaking resonance shifts around $0.84f_0$, due to the short-circuiting of the interleaved strips, while the antenna exhibits a very good impedance matching at $f_0$. As can be appreciated in Fig. 4 (d), where the total scattered power of the coated antenna evaluated at $f_0$ for different level of the antenna input power is reported, a smooth transition region from a cloaked to an uncloaked behavior appears from 0 *dBm* to 30 *dBm*, without any sharp variation. It is worth noticing that this power range is directly related to the characteristics of the loading diodes, and a careful choice of the component depending on the actual needs is, thus, required.

It is worth noticing that, the magnitude of the cloaking resonance at $0.84f_0$ is reduced in the HP scenario compared to the case of a perfect short circuit of the interleaved strips, due to the parasitic resistance $R_{HP}$. Moreover, the antenna is mismatched at the cloaking frequency $0.84f_0$, as expected from the fundamental cloaking limits discussed above. Finally, it is also worth remarking that the designed antenna system would have been always mismatched at $f_0$ if a linear conventional mantle cloak was exploited and, thus, it would have been an inefficient radiator for both LP and HP signals. On the contrary, the proposed non-linear metasurface cloak allows designing dipole antennas able to efficiently transmit high-power signals, while being invisible to low power ones.

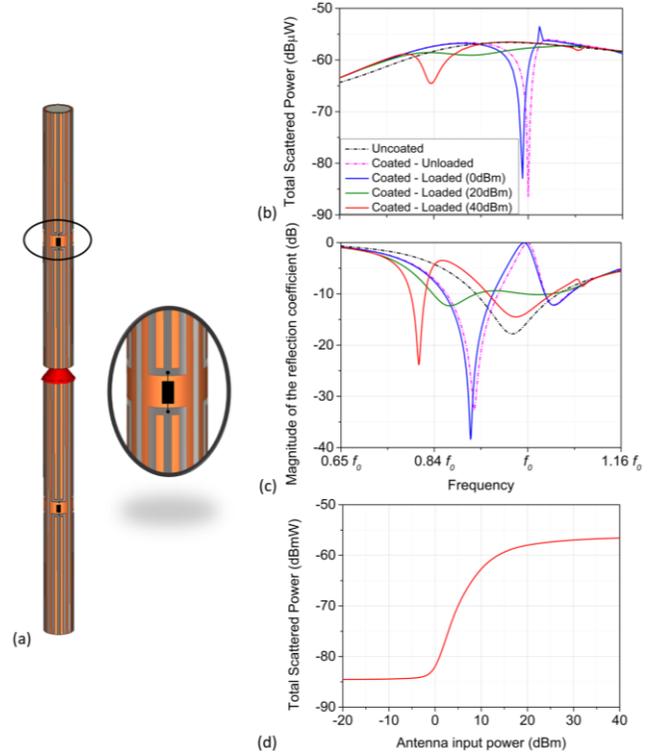

Fig. 4. (a) Dipole antenna coated by the non-linear mantle cloak. In the inset, we show the detail of the cloaking metasurface made of alternatively continuous and separated vertical metallic strips. Depending on the power level, the metallic strips are short/open-circuited by the loading diodes. (b) Total scattered power of the antenna for different scenarios: without the mantle cloak (uncoated case); with the diode-unloaded mantle cloak (coated-unloaded); with the diode-loaded mantle cloak for different values of the transmitted power (coated-loaded). (c) Magnitude of the antenna reflection coefficients for the same scenarios of (b). (d) Total scattered power evaluated at $f_0$ of the antenna coated by the non-linear mantle cloak for different values of the antenna input power.

## IV. NON-LINEAR CLOAKS FOR PHASED ARRAYS

In order to further investigate and demonstrate the possibilities offered by the proposed concept, a non-linear cloaking device has been also designed to enable unprecedented functionalities in antenna arrays. As it is well known, the scanning capabilities of phased arrays are usually limited by the presence of mutual coupling effects among the array elements [32]. Indeed, due to the mutual coupling between neighboring radiators, the embedded element factor of each radiator is usually non-omnidirectional, thus resulting in limited scanning ranges and undesired scan loss effects, even though the individual radiating element is omnidirectional when operating in free-space. To illustrate such a scenario, the case of a 3x3 array of half-wavelength dipoles is considered (see Fig. 5(a)). As expected, the embedded element factor of the central element is significantly different from the omnidirectional pattern of the isolated radiator because of the blockage effects induced by the presence of the surrounding elements [Fig. 5(b)].

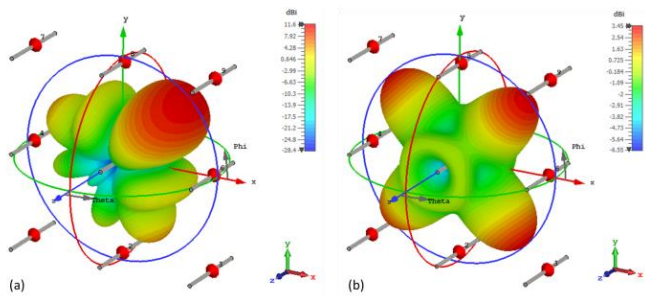

Fig. 5. (a) 3D radiation diagram of a phased array of half-wavelength dipole antennas, when the elements are excited to obtain the main beam pointing in a given direction. (b) 3D embedded element pattern of the central antenna when the external elements are closed on matched loads.

The possibility to circumvent this limitation is addressed hereinafter through the customization of a non-linear power-dependent mantle cloak. More specifically, a radiating system exhibiting a directive pattern when operating in the transmitting mode, while presenting an omnidirectional pattern in the receiving mode, is designed. This solution can be of particular interest for radar system able to scan selectively the environment when transmitting high-power pulses, and able to receive from all directions low-power scattered signals.

The considered configuration is depicted in Fig. 6(a) where all the elements of the 3x3 array, except the central one, are covered by power-dependent mantle cloaks. The central frequency of operation of the half-wavelength dipoles is $f_0 = 3$ GHz, and the array elements are coated by the inductive coating cloak discussed in the previous Section.

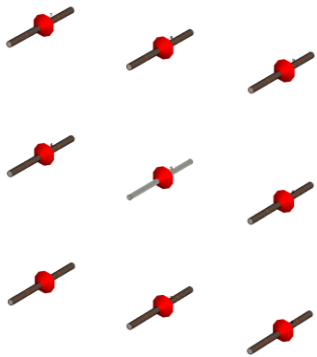

Fig. 6. Phased array made of half-wavelength dipole antennas. All the external elements are covered by the non-linear cloaking metasurfaces.

According to the above discussion, the array is expected to behave like a conventional phased array for HP signals (*i.e.*, in the transmitting mode). Conversely, the array does act as a half-wavelength dipole in the LP scenario (*i.e.*, in the receiving mode, when the surrounding elements are invisible to the central antenna thanks to the coating metasurfaces). This is confirmed by the 3D and 2D radiation diagrams reported in Fig. 7 for high and low power excitations. In particular, the embedded gain polar plots of the central element of the array are plotted in Fig. 7(c) for different scenarios. As it can be appreciated, the gains of the isolated and low-power excitation cases overlap. On the other hand, the curve in the high-power scenario matches the one obtained when the power-dependent cloaking devices are not present (uncoated case), meaning that the overall radiating system behaves as a conventional phased array. Thus, in this latter case, the radiation diagram is the one depicted in Fig. 7 (a).

It is worthwhile remarking that the discussed radiating system is only a benchmark example of the possibilities offered by the proposed approach. Further radiating systems with novel or enhanced functionalities could be envisioned by designing non-linear mantle cloaks and, thus, circumventing the inherent limitations of linear and passive cloaks for antennas that are made invisible at their own resonance frequencies.

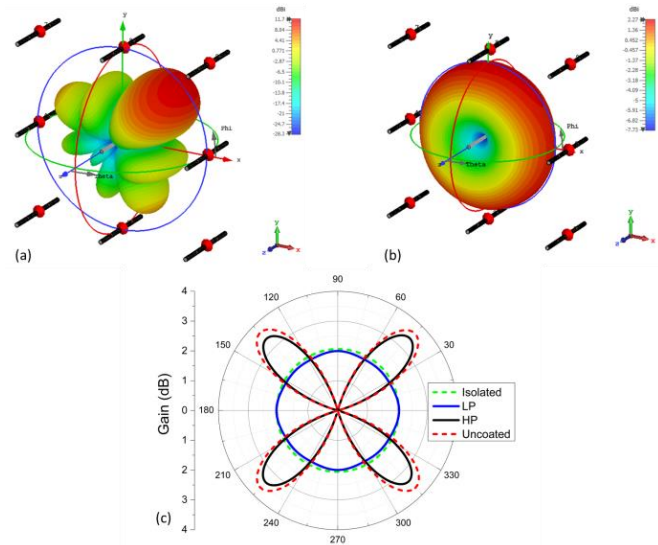

Fig. 7. (a) 3D radiation diagram of the coated phased array in the HP scenario. (b) 3D embedded element factor of the central antenna when the external elements are closed to matched loads. (c) Polar gain plots for different cases: external elements removed (isolated case); external elements without the cloaking devices (uncoated case); external elements with the cloaking devices in the low-power and high-power scenarios (LP and HP, respectively).

## V. CONCLUSIONS

An innovative methodological strategy to circumvent the fundamental bounds of mantle cloaking for antennas through the employment of non-linear cloaking metasurfaces has been introduced in this Communication. The proposed approach relies on the possibility to achieve a dynamic transformation of the geometry of the cloaking metasurface depending on the power-level of the impinging field. The presented idea has been numerically validated with full-wave co-simulations and realistic examples of dipole antennas covered by non-linear mantle cloaks, discussing also different applicative scenarios. In particular, an antenna application involving a phased array able to transmit towards an arbitrary pointing direction and receiving with an omnidirectional pattern has been presented. The illustrated results suggest the possibility to extend and generalize the proposed findings beyond the discussed scenarios to introduce new degrees of freedom for designing complex antenna systems in which both the electric and the radiation properties may be made selective with respect to the input power.